\documentclass[fleqn,twoside]{article}
\usepackage{espcrc2}

\usepackage{graphicx}
\usepackage[figuresright]{rotating}

\newcommand{\be}{\begin{equation}}
\newcommand{\ee}{\end{equation}}
\newcommand{\bea}{\begin{eqnarray}}
\newcommand{\eea}{\end{eqnarray}}
\newcommand{\Tr}{\mbox{Tr}}             
\newcommand{\inv}{\frac{1}}
\newcommand{\AmS}{{\protect\the\textfont2
  A\kern-.1667em\lower.5ex\hbox{M}\kern-.125emS}}

\title{Lattice Theories with Nonlinearly Realized Chiral Symmetry}

\author{S.~Chandrasekharan\address{Department of Physics, 
        Duke University, P.O.~Box 90305, Durham NC 27708-0305, USA},
        M.~Pepe\address[ITP_BE]{Institute for Theoretical Physics, 
        Bern University, Sidlerstrasse 5, CH-3012 Bern, Switzerland},
        F.~D.~Steffen\addressmark[ITP_BE]\thanks{Fellow within the 
        Postdoc-Programme of the DAAD.}
        and
        U.-J.~Wiese\addressmark[ITP_BE]\thanks{On leave from MIT.
        \newline
        Work supported by the DOE under grant DOE-FG02-96ER40945 
        and by the Schweizerischer Nationalfond.}}
       
\begin{document}

\begin{abstract}
  We present the lattice formulation of effective Lagrangians in which
  chiral symmetry is realized nonlinearly on the fermion fields. In
  this framework both the Wilson term removing unphysical doubler
  fermions and the fermion mass term do not break chiral symmetry.
  Our lattice formulation allows us to address non-perturbative
  questions in effective theories of baryons interacting with pions
  and in models involving constitutent quarks interacting with pions
  and gluons. With the presented methods, a system containing a
  non-zero density of static baryons interacting with pions can be
  studied on the lattice without encountering a complex action
  problem. This might lead to new insights into the phase diagram of
  strongly interacting matter at non-zero chemical potential.
  \vspace{1pc}
\end{abstract}

\maketitle

\section{INTRODUCTION AND OVERVIEW}

In order to realize chiral symmetry on the lattice, it is by now
understood that the lattice Dirac operator must satisfy the
Ginsparg-Wilson relation and not anticommute with $\gamma_5$ as
assumed by the Nielsen-Ninomiya theorem~\cite{Niedermayer:1998bi}. We
present an alternative view at chiral symmetry on the lattice within
low-energy effective theories (cf.~\cite{Chandrasekharan:2003wy} for
details).  Based on spontaneous chiral symmetry breaking, effective
Lagrangians can be constructed which involve an explicit pion field in
addition to phenomenological baryon~\cite{Weinberg:1966fm+X} or
constituent quark and gluon fields~\cite{Manohar:1983md}. The pion
field allows one to realize chiral symmetry nonlinearly on the fermion
fields. The fermion mass term then does not break chiral
symmetry~\cite{Weinberg:1966fm+X}.  We formulate such effective
theories on the lattice. The nonlinear realization of chiral symmetry
avoids the Nielsen-Ninomiya theorem and allows us to remove the
doubler fermions with a Wilson term while maintaining exact chiral
symmetry on the lattice. Our lattice formulation can be used to
address non-perturbative questions in effective theories, to study
static baryons at non-zero chemical potential, and to test the chiral
quark model~\cite{Manohar:1983md}.

\section{NONLINEAR REALIZATION OF CHIRAL SYMMETRY}

Let us illustrate the nonlinear realization of chiral symmetry for a
low-energy effective theory of pions and nucleons introduced
originally in~\cite{Weinberg:1966fm+X}.

\subsection{Continuum Formulation}

The chiral symmetry group of QCD in the limit of $N_f$ massless quark
flavors $G = SU(N_f)_L \otimes SU(N_f)_R \otimes U(1)_B$ is
spontaneously broken to the vector subgroup $H = SU(N_f)_{L=R} \otimes
U(1)_B$. Thus, massless Goldstone boson fields $U(x) \in G/H =
SU(N_f)$ govern the low-energy dynamics of QCD.  The corresponding
low-energy effective theory is given to lowest order in the Goldstone
boson momenta by the following
action~\cite{Weinberg:1978kz,Gasser:1983yg+X}:
\be
        S[U] = \int d^4x\ 
        \frac{F_\pi^2}{4} \Tr[\partial_\mu U^\dagger \partial_\mu U]
        \, ,
\label{Eq:S_U_continuum}
\ee
where $F_\pi$ is the pion decay constant. By construction, this action
is invariant under global chiral rotations, $L\otimes R \in SU(N_f)_L
\otimes SU(N_f)_R$, under which the Goldstone boson field transforms
as
\be
        U(x) \to U'(x) =  L U(x) R^\dagger \, .
\label{Eq:U^prime_cont}
\ee

To extend this theory to a low-energy effective theory for baryons, an
auxiliary field is defined:
\be
        u(x) = U(x)^{1/2} \, ,
\ee
which transforms such that~(\ref{Eq:U^prime_cont}) is satisfied:
\be
u(x) \to u(x)' = L u(x) V(x)^\dagger = V(x) u(x) R^\dagger \, .
\label{Eq:u^prime_cont}
\ee
The matrix $V(x) \in SU(N_f)$ is a local realization of the global
symmetry $G$,
\bea
V(x) & = & R\left[R^\dagger L U(x)\right]^{1/2} \big[U(x)^{1/2}\big]^\dagger 
\nonumber\\
& = & L\left[L^\dagger R U(x)^\dagger\right]^{1/2} U(x)^{1/2}
\, .
\eea
Treating $V(x)$ basically as a flavor gauge transformation, chiral
symmetry can be realized nonlinearly on the fermion fields.

Concentrating on the case $N_f=2$, a nucleon field---described by a
Dirac spinor $\Psi(x)$ and $\overline \Psi(x)$ in the fundamental
representation of $SU(N_f=2)$---transforms under the nonlinear
realization of chiral symmetry as follows
\bea
\Psi(x) & \to & \Psi(x)' = V(x) \Psi(x) \ ,
\\
\overline \Psi(x) & \to &  \overline \Psi(x)' = \overline \Psi(x) V(x)^\dagger \, .
\eea
A ``flavor covariant'' derivative can then be constructed with the
composite anti-Hermitean field
\be
v_\mu(x) = \inv{2}[u(x)^\dagger \partial_\mu u(x)+ u(x) \partial_\mu u(x)^\dagger]
\ee
transforming as a vector ``gauge'' field
\be
v_\mu(x) \to v_\mu'(x) = V(x)[v_\mu(x) + \partial_\mu] V(x)^\dagger
\, .
\ee
A composite Hermitean field can be build as well
\be
a_\mu(x) =  \frac{i}{2}[u(x)^\dagger\partial_\mu u(x)-u(x)\partial_\mu u(x)^\dagger] \, ,
\ee
which transforms as an axial vector
\be
a_\mu(x) \to a_\mu'(x) = V(x) a_\mu(x) V(x)^\dagger
\, . 
\ee

Now we can write down the leading terms of the Euclidean action
describing a chirally symmetric low-energy effective theory for
nucleons and pions
\bea
        S[U,\overline \Psi,\Psi] & = & S[U] 
        + \int d^4x \Big\{ \overline \Psi \gamma_\mu (\partial_\mu + v_\mu)\Psi
\nonumber\\
        && 
        +\, M \overline \Psi \Psi + i g_A \overline \Psi \gamma_\mu \gamma_5 a_\mu \Psi 
        \Big\}
        \, ,
\label{Eq:LEET_N_Pi_cont}
\eea
where $M$ is the nucleon mass generated by spontaneous chiral symmetry
breaking and $g_A$ is the coupling to the isovector axial current not
constrained by chiral symmetry. As advertized, the nucleon mass term
does not break chiral symmetry explicitly. This feature of the
nonlinear realization will allow us to perform lattice investigations
with massive fermions in the chiral limit.

\subsection{Lattice Formulation}

The contiuum discussion shows that the nonlinear realization of chiral
symmetry does not rely on the anticommutation of the Dirac operator
with $\gamma_5$. Therefore, the Nielsen-Ninomiya theorem does not
apply and we can use a Wilson term to remove the fermion doublers
without breaking chiral symmetry explicitly. The effective lattice
action of the low-energy theory for nucleons and pions can thus be
written in the following form:
\bea
&&
\hspace{-0.6cm}
S[U,\overline \Psi,\Psi] =
-\frac{F_\pi^2}{4}\sum_{x,\mu}\Tr[U_x^\dagger U_{x+\hat\mu}+U_{x+\hat\mu}^\dagger U_x]
\nonumber\\ && 
\hspace{-0.6cm}
+\,\inv{2}\sum_{x,\mu}(\overline\Psi_x\gamma_\mu V_{x,\mu}\Psi_{x+\hat\mu}-
\overline\Psi_{x+\hat\mu}\gamma_\mu V_{x,\mu}^\dagger\Psi_x)
\nonumber
\\ && 
\hspace{-0.6cm}
+\,\frac{i g_A}{2} \sum_{x,\mu}  
(\overline \Psi_x \gamma_\mu \gamma_5 A^L_{x,\mu}\! \Psi_{x}\! +\! 
\overline \Psi_{x+\hat\mu} \gamma_\mu \gamma_5 A^R_{x,\mu}\! \Psi_{x+\hat\mu})
\nonumber\\ && 
\hspace{-0.6cm}
+\,\frac{r}{2} \sum_{x,\mu} (2 \overline \Psi_x \Psi_x\! -\!
\overline \Psi_x V_{x,\mu} \Psi_{x+\hat\mu}\! -\!
\overline \Psi_{x+\hat\mu} V_{x,\mu}^\dagger \Psi_x)
\nonumber\\ && 
\hspace{-0.6cm}
+\,M \sum_x \overline \Psi_x \Psi_x
\label{Eq:LEET_N_Pi_latt}
\eea
with the Wilson term weighted by $r$. The lattice Goldstone boson
field $U_x$, its square root $u_x$, and the nucleon field $\Psi_x$ and
$\overline\Psi_x$ are obtained---together with the corresponding
transformation behavior---by naive discretization. The continuum
flavor ``gauge'' vector field $v_\mu(x)$ is realized by the flavor
``gauge'' parallel transporter
\bea
V_{x,\mu} &=& (\tilde V_{x,\mu} \tilde V_{x,\mu}^\dagger)^{-1/2}\,\tilde V_{x,\mu} \ , \\
\tilde V_{x,\mu} &=& \frac{1}{2}[u_x^\dagger u_{x+\hat\mu} + u_x u_{x+\hat\mu}^\dagger] \, .
\eea
which is an element of the $SU(N_f = 2)$ group even at finite lattice
spacing and transforms under lattice flavor ``gauge'' transformations
$V_x$ as
\be
V_{x,\mu} \to V_{x,\mu}' = V_x V_{x,\mu} V_{x+\hat\mu}^\dagger
\, 
\ee
%
The
continuum axial vector field $a_\mu(x)$ is realized by the lattice
fields $A^L_{x,\mu}$ and $A^R_{x,\mu}$ given
in~\cite{Chandrasekharan:2003wy}.

\section{APPLICATIONS}

The nonlinear realization of chiral symmetry on the
lattice---illustrated above only for the simplest case---allows a
non-perturbative description of flavor dynamics based on effective
Lagrangians. This opens up a new class of investigations.

\subsection{Low-Energy Effective Theories}

Within low-energy effective theories, we can study non-perturbative
phenomena on the lattice such as the spectrum of the $SU(N_f)$ chiral
rotor. While this can already be computed analytically in the
continuum~\cite{Leutwyler:1987ak}, lattice techniques will become
crucial in more complicated effective theories describing, for
example, nuclear matter.

\subsection{Static Baryons at Non-Zero \boldmath$\mu$}

Beyond the strict validity range of low-energy effective theories, the
presented methods can lead to new insights into the phase diagram of
strongly interacting matter at non-vanishing chemical potential.  In
the limit of static baryons, interesting simplifications arise which
we illustrate for $N_f =2$ in the continuum formulation.

A static nucleon, $M \rightarrow \infty$, at an undetermined position
$\vec x$ (with $g_A = 0$) is described by the spatial integral of the
Polyakov loop,
\be
\Phi[U] = \int d^3x \ \mbox{Tr}_F \ {\cal P} \exp[\int_0^\beta\!\!dt \ v_4(\vec x,t)] \ ,
\ee
where ${\cal P}$ denotes path ordering, $\beta = 1/T$ is the inverse
temperature, and $F$ refers to the trace in the fundamental
representation, which is real-valued for $SU(N_f=2)$. Then, taking
also the limit of an infinite chemical potential $\mu\to\infty$ such
that the difference $M-\mu$ remains finite, the grand canonical
partition function for a system of pions in the background of static
nucleons reduces to
\be
Z(\mu) = \!\!\int\!\!{\cal D}U \exp\{-S[U]+e^{-\beta(M-\mu)}\Phi[U]\} \ .
\ee
As $\Phi[U]$ takes on real values only, the phase diagram of this
model can be studied on the lattice without a complex action problem.
Taking into account non-zero quark masses and the associated small
explicit chiral symmetry breaking, the dependence of the phase
structure on the quark masses can be determined as well. Since the
static nucleon model shares all global symmetries with QCD, such
investigations might help to extract the universal properties of the
QCD phase diagram. To investigate the dependence on the strange quark
mass, the model can be generalized to $SU(N_f = 3)$. Again no complex
action problem arises since now the baryons transform in the adjoint
representation.

\subsection{Constituent Quarks on the Lattice}

A nonlinear realization of chiral symmetry was used also in the chiral
quark model of Georgi and Manohar~\cite{Manohar:1983md}. This model is
obtained from~(\ref{Eq:LEET_N_Pi_cont}) by replacing the nucleons with
colored constituent quarks and by adding gluon dynamics. If this model
provides an appropriate low-energy description for QCD, one could
understand why the non-relativistic quark model works. Our lattice
formulation~\cite{Chandrasekharan:2003wy} provides the ideal means to
clarify this issue as there are open non-perturbative questions
concerning, for example, the confinement of constituent quarks and the
assumption of the chiral symmtry breaking scale $\Lambda_{\chi SB}$
being larger than the confinement scale $\Lambda_{\mathrm{QCD}}$.

\section{OUTLOOK AND CONCLUSION}

Is it interesting to ask if nonlinearly realized chiral symmetry can
lead to a new approach to lattice QCD beyond low-energy effective
theories. In order to answer this question, one may attempt to relate
lattice QCD with Ginsparg-Wilson fermions to a lattice theory with
explicit pion fields and a nonlineary realized chiral symmetry (cf.\ 
also~\cite{Brower:lattice2003}). In any case, the nonlinearly realized
chiral symmetry provides an alternative perspective at chiral symmetry
on the lattice.

\end{document}